\documentclass[prl,a4paper,showpacs,superscriptaddress,twocolumn]{revtex4-1}
\usepackage{amsmath}
\usepackage{amssymb}
\usepackage{amsfonts}
\usepackage{graphicx}
\usepackage{color}
\usepackage{epstopdf}
\usepackage{bm}

\DeclareMathOperator{\Tr}{Tr}

\DeclareMathOperator{\Prob}{Prob}
 \DeclareMathOperator{\Imag}{Im}
\DeclareMathOperator{\Real}{Re}

\DeclareMathOperator{\erf}{erf}

\newcommand{\bb}{\begin{equation}}
\newcommand{\ee}{\end{equation}}

\begin{document}

\title{Statistics of Spin Fluctuations in Quantum Dots with Ising Exchange}

\author{D.S. Lyubshin} 
\affiliation{L.D. Landau Institute for Theoretical Physics RAS,
Kosygina street 2, 119334 Moscow, Russia}
\affiliation{Moscow
Institute of Physics and Technology, 141700 Moscow, Russia}
\author{A.U. Sharafutdinov}
\affiliation{L.D. Landau Institute for Theoretical Physics RAS,
Kosygina street 2, 119334 Moscow, Russia}
\author{I.S. Burmistrov}
\affiliation{L.D. Landau Institute for Theoretical Physics RAS,
Kosygina street 2, 119334 Moscow, Russia}
\affiliation{Moscow
Institute of Physics and Technology, 141700 Moscow, Russia}

\date{\today}

\begin{abstract}
We explore the effect of single-particle level fluctuations on the Stoner instability in a QD with a strong spin-orbit coupling in the framework of the universal Hamiltonian with the Ising exchange interaction. We reduce the problem to studying the statistics of extrema of a certain Gaussian process and demonstrate that, in spite of the randomness of the single-particle levels, the longitudinal spin susceptibility and all its moments diverge simultaneously at the point of the Stoner instability which is determined by the standard criterion involving the mean level spacing only.
\end{abstract}


\maketitle


{\it Introduction.---}%
In the last two decades the physics of quantum dots (QDs) attracted a lot of interest from both experimentalists and theorists~\cite{Alhassid2000,Wiel,ABG,Hanson,Ullmo2008}. Under the assumption $E_{\rm Th}/\delta \gg 1$, where
$E_{\rm Th}$ and $\delta$ denote the Thouless energy and the mean single-particle level spacing respectively, an effective zero-dimensional Hamiltonian has been derived \cite{KAA}. This so-called universal Hamiltonian (UH) provides a convenient framework for the theoretical description of QDs. In the UH the electron-electron interaction that involves a set of matrix elements in the single-particle basis is reduced to just three parameters: the charging energy ($E_c$), the ferromagnetic exchange ($J>0$) and the interaction in the Cooper channel.

At low temperatures $T\ll E_c$ the additional cost due to the charging energy restricts the probability of real electron tunneling through a QD, a phenomenon known as the Coulomb blockade \cite{CB}. This leads to suppression of the tunneling density of states (TDOS) in QDs for $T\ll E_c$ \cite{KamenevGefen1996,SeldmayrLY}. Although typically $E_c\gg \delta$, a small enough exchange interaction $J\lesssim \delta/2$ is important for a quantitative description of the experiments on transport through QDs at $T\lesssim \delta$ in a two-dimensional electron gas \cite{QDLowT}. The exchange interaction of a large QD can be estimated by the Fermi-liquid interaction parameter ($F_0^\sigma$) of the corresponding bulk material: $J/\delta = -F_0^\sigma$. As well-known for bulk materials, strong enough exchange interaction leads to a Stoner instability at $F_0^\sigma=-1$ and a corresponding quantum phase transition between a paramagnet and a ferromagnet. In QDs an intermediate case of the ground state (GS) having a finite value of spin can be realized for $\delta/2 \lesssim J<\delta$ in the case of  the equidistant single-particle spectrum \cite{KAA}. As $J$ increases towards $\delta$, the GS spin increases and at $J=\delta$ all electrons in a QD become spin polarized. This phenomenon of mesoscopic Stoner instability disappears in the thermodynamic limit $\delta \to 0$. Due to the entanglement of the charge and spin degrees of freedom in the UH, the mesoscopic Stoner instability affects the electron transport through a QD. For example, it leads to an additional nonmonotonicity of the energy dependence of the TDOS \cite{KiselevGefen,BGK1,BGK2} and to the enhancement of the shot noise \cite{Koenig2012}. The Cooper channel interaction in the UH describes the superconducting correlations in QDs \cite{SCinQD}. We shall assume that there is no attraction in the Cooper channel and, therefore, disregard it below \cite{KAA}. We also neglect the corrections to the UH due to the fluctuations in the matrix elements of the interaction \cite{Altshuler1997,Mirlin1997}, which are small in the regime $\delta/E_{\rm Th}\ll 1$ but can lead to interesting physics beyond the UH \cite{Ullmo2008}.

In the presence of a spin-orbit coupling the UH description of a QD breaks down. For a large spin-orbit length (weak spin-orbit coupling), $\lambda_{SO} \gg L$, where $L$ is a typical size of the QD, fluctuations of the matrix elements of the interaction cannot be neglected \cite{AlhassidSO,SOinQD}. However, for a QD fabricated in a two-dimensional electron gas the orbital degrees of freedom are coupled to in-plane components of the spin only. Thus in the regime $(\lambda_{SO}/L)^2 \gg (E_{\rm Th}/\delta)(L/\lambda_{SO})^4\gg 1$   the UH description is restored but with the Ising exchange interaction \cite{AlhassidSO,AF2001}.
Contrary to case of the Heisenberg exchange, there is no mesoscopic Stoner instability within the UH with the Ising exchange ($J_z>0$) for the equidistant single-particle spectrum \cite{KAA}. As a consequence, the TDOS is almost independent of $J_z$ while the longitudinal spin susceptibility $\chi_{zz}$ is independent of $T$ as in a clean Fermi liquid \cite{KiselevGefen,Boaz}.

In the interacting electron systems a disorder-induced finite temperature transition between the paramagnetic and the ferromagnetic phases is possible in low dimensions $d\leqslant 2$ \cite{Finkelstein,KamenevAndreev1998}. In $d=3$ the Stoner instability can be shifted towards the smaller values of the exchange interaction due to disorder \cite{Nayak2005}. In the UH description the disorder is translated into randomness of the single-particle levels. The latter is crucial in the case of the Ising exchange since the average $\chi_{zz}$ acquires a $T$-dependent contribution of Curie type due to the level fluctuations \cite{KAA}. The Curie-type contribution dominates at low enough $T$ and for $\delta-J_z\ll\delta$. In this regime the level fluctuations become strong with respect to the distance $\delta-J_z$ to the average position of the Stoner instability at $J_z=\delta$ (albeit small when compared to the temperature). This implies that, although for $\delta-J_z\ll\delta$ the QD is in the paramagnetic phase on average, for a particular realization of the single-particle levels the QD can be fully spin-polarized. Such events should affect the tail of the distribution function for $\chi_{zz}$, but how exactly? Can it be possible that at $T=0$ the level fluctuations shift the position of the Stoner instability from $J_z=\delta$ and lead to the existence of a finite temperature transition between the paramagnetic and the ferromagnetic phases?

In this Letter we address these questions within the UH with the Ising exchange interaction approach. We demonstrate that for $\delta-J_z\ll\delta$ the statistical properties of the spin susceptibility in the temperature range $\delta\ll T \ll \delta J_z/(\delta-J_z)$ are determined by the statistics of the extrema of a certain Gaussian process with drift that locally resembles a fractional Brownian motion (FBM) with Hurst exponent $H=1-\epsilon$ with $\epsilon \to 0$ (recall that the FBM with Hurst exponent $H$ is the Gaussian process $B_H(t)$ with zero mean and the two-point correlation function $\overline{[B_H(t)-B_H(t^\prime)]^2} = |t-t^\prime|^{2H}$). We estimate the complementary cumulative distribution function (CCDF) for $\chi_{zz}$ and show that all moments of $\chi_{zz}$ are finite for $J_z<\delta$. Thus our results mean that at $T\gg\delta$ in spite of the presence of strong level fluctuations the Stoner instability still occurs at $J_z=\delta$.


{\it The partition function.---} We consider the following universal Hamiltonian with direct Coulomb and Ising exchange interactions \cite{KAA}:
\begin{equation}
H =\sum_{\alpha,\sigma}\epsilon_\alpha a^{\dag}_{\alpha\sigma}a_{\alpha\sigma}+E_c(\hat{N}-N_0)^2 -J_z\hat{S}_z^2 .
\label{eq:UH:start}
\end{equation}
For an isolated lateral QD with Rashba and Dresselhaus spin-orbit couplings the statistics of single-particle energies $\epsilon_\alpha$ is described by the unitary Wigner-Dyson ensemble (class A) \cite{AF2001}. The operators of the total number of particles $\hat{N}=\sum_{\alpha,\sigma}a^{\dag}_{\alpha,\sigma}a_{\alpha,\sigma}$ and the total spin $\hat{\bm{S}}=(1/2)\sum_{\alpha\sigma\sigma^\prime} a^\dag_{\alpha\sigma}\bm{\sigma}_{\sigma\sigma^\prime}a_{\alpha\sigma^\prime}$ are given as usual in terms of the single-particle creation ($a^{\dag}_{\alpha,\sigma}$) and annihilation ($a_{\alpha,\sigma}$) operators and the Pauli matrices $\bm{\sigma}$.

Since the operators for the number of spin-up and spin-down electrons commute with $H$, the grand partition function $Z= \Tr e^{-\beta H+\beta \mu \hat N }$ can be written as \cite{Boaz}
\begin{equation}
Z= \sum\limits_{n_\uparrow,n_\downarrow} Z_{n_\uparrow}Z_{n_\downarrow}
e^{-\beta E_c(n-N_0)^2+\beta J_z m^2+\beta\mu n}.
\label{eqZb1}
\end{equation}
The integers $n_\uparrow$ and $n_\downarrow$ represent the number of spin-up and spin-down electrons respectively. The total number of electrons is $n = n_\uparrow+n_\downarrow$, and $m =
(n_\uparrow - n_\downarrow)/2$ is the value of $S_z$. The factor $Z_{n_\uparrow}$ ($Z_{n_\downarrow}$) is the canonical partition function for $n_\uparrow$ ($n_\downarrow$) noninteracting spinless electrons. They take into account the contributions due to the single-particle energies and are given by the Darwin-Fowler integral: $Z_n=\int_0^{2\pi} \frac{d\theta}{2\pi}\, e^{-i n \theta}\prod_{\gamma}\left (1+ e^{i\theta-\beta\epsilon_\gamma}\right )$.
There is a convenient integral representation for $Z$ which is exactly equivalent to Eq. \eqref{eqZb1}:
\begin{gather}
Z= \sum_{k\in \mathbb{Z}}e^{-\beta E_c(k-N_0)^2} \int_{-\pi T}^{\pi T}\frac{d\phi_0}{2\pi T} \,  e^{i\beta \phi_0 k} \widetilde{Z}(\mu-i\phi_0) ,
 \notag \\ \widetilde{Z}(\mu) =  \int_{-\infty}^{\infty} \frac{dh}{\sqrt{\pi \beta J_z}}\,   e^{-\frac{h^2}{\beta J_z}}
\prod_{\sigma}e^{-\beta \Omega_0(\mu+h \sigma/\beta)} .
\label{IR}
\end{gather}
Here $\Omega_0(\mu)=-T \ln \prod_{\gamma}\left (1+e^{-\beta (\epsilon_{\gamma}-\mu)}\right )$ stands for the thermodynamic potential of free spinless electrons; $\phi_0$ and $h$ are the zero-frequency Matsubara components of the electric potential and the magnetic field that can be used to decouple the direct Coulomb \cite{KamenevGefen1996} and exchange interaction \cite{Boaz,Saha2012} terms respectively.

At not very low temperatures $T\gg \delta$ we can perform integration over $\phi_0$ in Eq. \eqref{IR} in the saddle-point approximation \cite{KamenevGefen1996,EfetovTscherisch}. In this case the grand canonical partition function  factorizes: $Z = Z_C Z_S$, where
\begin{equation}
Z_C = \sqrt{\frac{\beta\Delta}{4\pi}} \sum_{k\in\mathbb{Z}} e^{-\beta E_c(k-N_0)^2 +\beta (\tilde{\mu}-\mu) k -2 \beta \Omega_0(\tilde\mu)} , \label{ZC1}
\end{equation}
describes the effect of the charging energy. The exchange interaction is encoded in $Z_S = \exp[2\beta\Omega_0(\tilde{\mu})]\widetilde{Z}(\tilde{\mu})$. Here $\tilde\mu$ is the solution of the saddle-point equation: $N_0 = -2 \partial \Omega_0(\tilde{\mu})/\partial \tilde{\mu}$, and $\Delta^{-1} = - \partial^2\Omega_0(\tilde{\mu})/\partial \tilde{\mu}^2$ stands for the thermodynamic density of states for the chemical potential $\tilde{\mu}$. The thermodynamic potential $\Omega_0(\tilde{\mu})$ depends on a particular realization of the single-particle spectrum via the single-particle density of states $\nu_0(E)=\sum_\alpha \delta(E+\tilde\mu-\epsilon_\alpha)$. Provided $h^2\ll \exp(\beta\tilde{\mu})$, we find
\begin{gather}
\beta\sum_\sigma \Bigl [\Omega_0(\tilde{\mu})-\Omega_0(\tilde{\mu}+ h\sigma/\beta)\Bigr ] = \frac{h^2}{\beta \delta}-V(h) ,\notag \\
V(h) = -\int_{-\infty}^\infty dE\, \delta\nu_0(E)\,  \ln \left [ 1+\frac{\sinh^2(h/2)}{\cosh^2(\beta E/2)}\right ] ,\label{Vh_Def}
\end{gather}
where $\delta \nu_0(E)$ stands for the deviation of the density of states $\nu_0(E)$ from its average value $1/\delta\equiv \overline{1/\Delta}$.

The longitudinal spin susceptibility is fully determined by the partition function, $\chi_{zz} = \partial \ln Z/\partial J_z$. Since $Z_C$ is independent of $J_z$ and therefore does not affect the spin susceptibility, we will discuss $Z_S$ only. We note that the normalization is such that $Z_S=1$ for $J_z=0$. According to Eq. \eqref{eqZb1}, $Z$ increases with $J_z$; it follows that $Z_S\geqslant 1$. It is useful to write down $Z_S$ explicitly:
\begin{equation}
Z_S = \sqrt{\bar{J}_z/J_z}
 \int_{-\infty}^\infty \frac{dh}{\sqrt{\pi}}
\exp \left [ -h^2-V\left (h\sqrt{\beta \bar{J}_z}\right ) \right ] ,
\label{eq:ZS:Xi}
\end{equation}
where $\bar{J}_z = \delta J_z/(\delta-J_z)$ is the renormalized exchange interaction. For the equidistant spectrum ($V=0$), Eq. \eqref{eq:ZS:Xi} yields $Z_S=\sqrt{\bar{J}_z/J_z}$ and $\chi_{zz}=1/[2(\delta-J_z)]$.


{\it Level fluctuations.---}Although the density of states $\nu_0(E)$ has non-Gaussian statistics, $V(h)$ is an even in $h$ Gaussian random function for $\max\{|h|,T/\delta\}\gg 1$~\cite{Mehta}. It has zero mean and the following two-point correlation function (see the Supplemental Material \cite{EPAPS}):
\begin{gather}
\overline{V(h_1)V(h_2)}
= \sum_{\sigma=\pm} L(h_1+\sigma h_2)- 2 L(h_1) - 2L(h_2) ,  \notag \\
L(h) = \frac{2}{\pi^2\bm{\beta}} \int_0^{|h|} dt \, t\,  \left [\Real \psi\left (1+\frac{i t}{2\pi}\right )+\gamma\right ] .
\label{corrVV}
\end{gather}
Here  $\psi(z)$ is the Euler digamma function, $\gamma=-\psi(1)$ is the Euler--Mascheroni constant, and $\bm{\beta}=2$ since the energy levels $\epsilon_\alpha$  in Eq. \eqref{eq:UH:start} are described  by the unitary Wigner-Dyson ensemble. The asymptotics of $L(h)$ are as follows~\cite{BGK1,EPAPS}:
\begin{equation}
L(h) = \frac{h^2}{\pi^2\bm{\beta}}\begin{cases}
\zeta(3) h^2/(8\pi^2),\quad & |h|\ll 1 ,\\
\ln[|h|/(2\pi)]+\gamma-1/2, \quad & |h|\gg 1 .
\end{cases} \label{Sassymp}
\end{equation}


{\it Average spin susceptibility.---} As well known, the level fluctuations are small at $T\gg \delta$. In this regime one can find from Eq.~\eqref{Sassymp} that the variance of the level spacing is $\overline{(\Delta-\delta)^2}/\delta^2 = 3\zeta(3) \delta^2/(2\pi^4\bm{\beta}T^2) \ll 1$ \cite{BGK1}. Therefore it seems that in order to find the average spin susceptibility at $T\gg \delta$ it is enough to substitute $1/\Delta$ for $1/\delta$ in the expression for $\chi_{zz}$ obtained above for the equidistant spectrum. Performing the expansion to the second order in $\Delta-\delta$, we find the average spin susceptibility
\begin{equation}
\overline{\chi}_{zz} = \frac{1}{2(\delta-J_z)} \left [1  + \frac{3\zeta(3)}{2\pi^4 \bm{\beta}}  \frac{\delta \bar{J}^2_z}{J_z T^2} \right ].
\label{eq:chizz:regI}
\end{equation}
This result indicates that the effect of fluctuations is small only at temperatures $T\gg \bar{J}_z$. For such temperatures the fluctuations of the level spacing are small in comparison with the distance to the average position of the Stoner instability, $\overline{(\Delta-\delta)^2} \ll (\delta-J_z)^2$. If $\delta-J_z\ll \delta$, the renormalized exchange interaction $\bar{J}_z\gg \delta$ and there is a wide interval of temperatures $\bar{J}_z\gg T \gg \delta$ where the effect of level fluctuations can be strong. We stress that the dependence of $\overline{\chi}_{zz}$ on $T$ appears only due to level fluctuations. The result~\eqref{eq:chizz:regI} can be also obtained from Eq.~\eqref{eq:ZS:Xi} by means of the second order perturbation theory in $V$ using the asymptotic expression \eqref{Sassymp} at $|h|\ll 1$.

For temperatures $T\ll \bar{J}_z$ the integral in the r.h.s. of Eq. \eqref{eq:ZS:Xi} is dominated by $|h| \sim \sqrt{\bar{J}_z/T} \gg 1$. Hence in this case to evaluate the integrals involved in the perturbation theory in $V$ one has to use the asymptotic formula \eqref{Sassymp} at $|h|\gg 1$.  Expanding Eq.~\eqref{eq:ZS:Xi} to the fourth order in $V$ and performing the averaging of $\ln Z_S$ with the help of Eqs.~\eqref{corrVV} and \eqref{Sassymp} we obtain \cite{EPAPS}
\begin{equation}
\overline{\chi}_{zz} = \frac{1}{2(\delta-J_z)} \left [ 1 + \frac{\bar{J}_z \ln 2}{\bm{\beta}\pi^2 T}+a_2 \left (\frac{\bar{J}_z}{\bm{\beta}\pi^2 T}\right )^2 \right ] ,
\label{eq:chizz:regII}
\end{equation}
where $a_2 \approx 0.29$. From Eq.~\eqref{eq:chizz:regII} we see that the perturbation theory in $V$ is justified only for $T\gg \bar{J}_z/(\pi^2\bm{\beta})$. Therefore the result~\eqref{eq:chizz:regII} is valid in the range $\bar{J}_z \gg T\gg \bar{J}_z/(\pi^2\bm{\beta})$. In this regime the fluctuations of the spin susceptibility around its average value are small, $\overline{(\chi_{zz}-\overline{\chi}_{zz})^2}/(\overline{\chi}_{zz})^2 \propto \bar{J_z}/(\pi^2 \bm{\beta} T) \ll 1$.

\begin{figure}[t]
\centerline{\includegraphics[width=8cm]{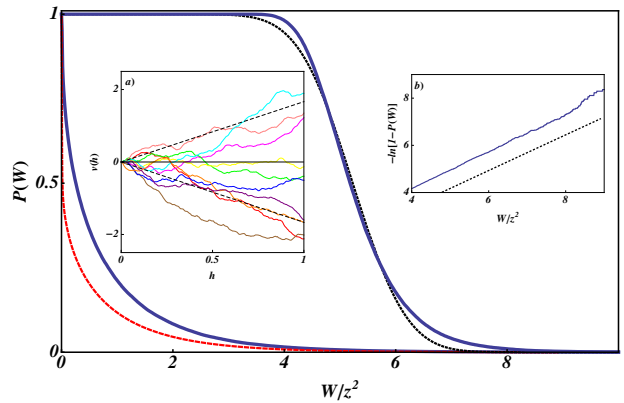}}
\caption{(Color online) The dependence of $\mathcal{P}(W)$ on $W/z^2$ at $T=3\delta$ obtained numerically for $J_z/\delta= 0.94$ ($z\approx 0.5$) (upper solid curve) and $J_z/\delta=
0.99994$ ($z\approx 16.8$) (lower solid curve). The black dotted curve is the CCDF for the normal distribution with mean and variance as one can find from the lowest order perturbation theory in $V$ for $T=3\delta$ and $J_z/\delta= 0.94$ \cite{EPAPS}. The red dashed curve is the CCDF of the degenerate process $\tilde{v}(h)$ for $T=3\delta$ and $J_z/\delta= 0.99994$ \cite{EPAPS}. Inset a): several realizations of the process $v(h)$; dashed lines $\pm 2 h \sqrt{\ln 2}$ are guides for an eye. Inset b): Comparison of the tail of $\mathcal{P}(W)$ obtained numerically for $J_z/\delta=
0.99994$ ($z\approx 16.8$) and asymptotic result \eqref{eq:Prob:exp}. }
\label{Fig:Process}
\end{figure}

{\it Tail distribution for $\ln Z_S$.---} The perturbative result~\eqref{eq:chizz:regII} suggests that the spin susceptibility can be strongly affected by level fluctuations at $\bar{J}_z/(\pi^2 \bm{\beta}) \gg T \gg \delta$. Such regime is realized in the close vicinity of the average position of the Stoner instability $\delta-J_z\ll  \delta/(\pi^2 \bm{\beta})$. However, if the effect of level fluctuations is strong then it is useful to know not only the average spin susceptibility but also all its moments. With this in mind, we investigate the CCDF for $\ln Z_S$ at temperatures in the interval $\bar{J}_z/(\pi^2 \bm{\beta}) \gg T \gg  \delta$. In this range of temperatures, the integral in the r.h.s. of Eq.~\eqref{eq:ZS:Xi} is dominated by the large values of $|h|$. Using asymptotic expression~\eqref{Sassymp}, one can check that for $|h_{1}|,|h_2|\gg 1$ the two-point correlation function \eqref{corrVV} is homogeneous of degree two: $\overline{V(u h_1)V(u h_2)} = u^2 \overline{V(h_1) V(h_2)}$~\cite{KAA}. We can therefore substitute $z v(h)$ for the random function $V(h\sqrt{\beta \bar{J}_z})$ where $z=\sqrt{\beta \bar{J}_z/(\pi^2 \bm{\beta})}$; the Gaussian random process $v(h)$ has zero mean, possesses the property $v(h)=v(-h)$, and its correlation function is given by
\begin{align}
\overline{v(h_1)v(h_2)} & =
\frac{1}{2}\sum_{\sigma=\pm} (h_1+\sigma h_2)^2 \ln (h_1+\sigma h_2)^2 \notag \\
& - h_1^2\ln h_1^2 -h_2^2 \ln h_2^2 .
\label{eq:vvh:def}
\end{align}
Since $\overline{\bigl [ v(h+u)-v(h)\bigr ]^2} = - 2 u^2 \ln |u| + O(u^2)=O(u^{2H})$ for any $H=1-\epsilon<1$, the trajectories of $v(h)$ are continuous and its increments are strongly positively correlated (see inset in Fig.~\ref{Fig:Process}). In fact the process $v(h)$ is in many aspects close to the ballistic case $\tilde v(h)=\xi |h|$, where $\xi$ is a Gaussian random variable (recall that $\tilde v(h)$ is the unique process with $H=1$). We mention that the process $v(h)$ has arisen before in a seemingly unrelated context~\cite{BGT}.

The average moments of $\ln Z_S$ can be conveniently written as $\overline{[\ln Z_S]^k}= k \int_{0}^{\infty}dW W^{k-1} \mathcal{P}(W)$ where the function $\mathcal{P}(W)$ is the complementary cumulative distribution function, i.e. the probability for $\ln Z_S$ to exceed $W$: $\mathcal{P}(W)\equiv \Prob\{\ln Z_S> W\}$. We note that $\mathcal{P}(0)=1$, $\mathcal{P}(\infty)=0$ and $\mathcal{P}(W)$ is monotonously decreasing.  Although we cannot find a closed analytical expression for $\mathcal{P}(W)$, we bound it from above to prove that all moments of $\ln Z_S$ (and consequently all moments of $\chi_{zz}$) are finite for $J_z<\delta$. We first split the Gaussian weight $\exp(-h^2)$ in the integral in the r.h.s. of Eq.~\eqref{eq:ZS:Xi} and obtain ($0<\gamma<1$ is an arbitrary splitting parameter)
\begin{equation}
Z_S \leqslant \frac{2\sqrt{\bar{J}_z}}{\sqrt{\pi \gamma J_z}} \int\limits_{0}^{\infty} dh \, e^{-\frac{(1-\gamma) h^2}{\gamma}}   \max\limits_{h\geqslant 0} \Bigl \{e^{-h^2-\frac{z v(h)}{\sqrt{\gamma}}}\Bigr \} .
\label{eq:split}
\end{equation}
The inequality \eqref{eq:split} allows us to reduce the problem of finding an upper bound for $\mathcal{P}(W)$ to studying the statistics of the maxima of the Gaussian process $Y(h)=-h^2-(z/\sqrt{\gamma})v(h)$ which locally resembles a FBM with drift. Indeed, from Eq. \eqref{eq:split} we find
\begin{equation}
\mathcal{P}(W) \leqslant \Prob \Bigl \{ \max\limits_{h\geqslant 0} Y(h) > W + \frac{1}{2}\ln \frac{(1-\gamma) J_z}{\bar{J}_z} \Bigr \} .
\end{equation}
To give an upper bound for the probability $\Prob\{ \max\limits_{h\geqslant 0} Y(h) > w\}$ we employ an auxiliary Gaussian process $X(h) = -h^2+(2 z\sqrt{\ln 2}/\sqrt{\gamma})B(h^2)$ where $B(h)$ is the standard Brownian motion (the Hurst exponent $H=1/2$). The processes $Y(h)$ and $X(h)$ satisfy the conditions for the Slepian's inequality \cite{Slepian},
$\Prob\{ \max\limits_{h\geqslant 0} Y(h) > w\}\leqslant \Prob\{\max\limits_{h\geqslant 0} X(h) > w\}$.
Ruin probabilities for $X(h)$ are trivial to compute \cite{Asmussen}, and we find the following upper bound for the CCDF \cite{EPAPS}:
\begin{equation}
\mathcal{P}(W) \leqslant  \exp\left\{-\frac{\gamma}{2z^2\ln 2}\left [W+\frac{1}{2}\ln \frac{(1-\gamma) J_z}{\bar{J}_z}\right ] \right\} .
\label{eq:tail}
\end{equation}
From Eq. \eqref{eq:tail} it follows that all moments of $\ln Z_S$ (and hence all moments of $\chi_{zz}$) are finite for $J_z<\delta$ for temperatures in the range $\bar{J}_z/(\pi^2\bm{\beta})\gg T\gg \delta$. Therefore even in the presence of the strong level fluctuations the Stoner instability occurs at $J_z=\delta$. For $J_z<\delta$ and for temperatures $T\gg \delta$ the QD is in the paramagnetic state.

For $z\gg 1$ the saddle-point approximation in Eq. \eqref{eq:ZS:Xi} becomes exact and the statistics of $\ln Z_S$ reduces to the statistics of maxima of $Y(h)$ directly. Since local behavior of $Y(h)$ may be compared to that of the FBM with Hurst exponent $H=1-\epsilon$, one can adapt the results of Ref. \cite{HuslerPiterbarg} for locally stationary processes and find that for $W\gg 2 z^2 \ln 2$ with logarithmic accuracy \cite{EPAPS}
\begin{equation}
\mathcal{P}(W) \propto \left ( \frac{z^2}{W} \ln \frac{W}{z^2} \right)^{1/2}\exp \left ( - \frac{W}{2 z^2 \ln 2} \right ).
\label{eq:Prob:exp}
\end{equation}
This result valid in the temperature range $\bar{J}_z/(\pi^2\bm{\beta})\gg T\gg \delta$ is consistent with the upper bound \eqref{eq:tail}. To illustrate the result \eqref{eq:Prob:exp} we approximate the Gaussian process $v(h)$ by a degenerate one $\tilde{v}(h) = \xi |h|$ where $\xi$ is the Gaussian random variable with zero mean $\overline{\xi}=0$ and variance $\overline{\xi^2}=4\ln 2$. We estimate the partition function \eqref{eq:ZS:Xi} as $Z_S \simeq \sqrt{\bar{J}_z/J_z} \exp(z^2 \xi^2/4) \Bigl [1 - \erf(z \xi/2) \Bigr ]$. The large values of $Z_S$ correspond to large negative values of $\xi$ such that $\ln Z_S \approx z^2 \xi^2/4$. Hence we find that for $z\gg 1$ the tail of the function $\mathcal{P}(W)$ is given by Eq.~\eqref{eq:Prob:exp} without the logarithm in the pre-exponent \cite{EPAPS}.  As shown in Fig. \ref{Fig:Process} the overall behavior of $\mathcal{P}(W)$ for $z\gg 1$ is well enough approximated by the CCDF for the degenerate process $\tilde{v}(h)$. Also  we mention that the behavior of $\mathcal{P}(W)$ for $z\gg 1$ is very different from its behavior at $z\lesssim 1$. For the later, $\mathcal{P}(W)$ is given by the CCDF for the normal distribution (see Fig. \ref{Fig:Process}).

{\it Average moments of $\chi_{zz}$.---} Equation~\eqref{eq:Prob:exp} implies that the average moments of $\ln Z_S$ scale as $\overline{(\ln Z_S)^k} \sim z^{2k}$ for $z\gg 1$. Hence for $\delta \ll T \ll \bar{J}_z/(\pi^2\bm{\beta})$ the $k$-th moment of the spin susceptibility is given by
\begin{equation}
\overline{\chi_{zz}^k} \propto \left [\frac{\delta^2}{\pi^2 \bm{\beta} (\delta-J_z)^2 T}\right ]^k  , \quad k=1, 2, \dots .
\label{eq:chi:kth}
\end{equation}
The result \eqref{eq:chi:kth} can be obtained from the saddle-point analysis of the integral in the r.h.s. of Eq. \eqref{eq:ZS:Xi}, i.e., in essence, by Larkin-Imry-Ma type arguments~\cite{Larkin,ImryMa}. The scaling of the average spin susceptibility (Eq. \eqref{eq:chi:kth} with $k=1$) was indeed found in Ref. \cite{KAA} using arguments of Larkin-Imry-Ma type.

{\it Summary.---} To summarize, we have studied the Stoner instability in a QD within the UH with the Ising exchange interaction. We demonstrated that in the regime $\delta-J_z\ll\delta$ where the level fluctuations are dangerous all moments of the spin susceptibility $\chi_{zz}(T)$ are finite at temperatures $T\gg \delta$ for $J_z<\delta$. This means that i) the Stoner instability is not shifted by the level fluctuations away from $J_z=\delta$ and ii) randomness in the single-particle levels does not lead to transition at finite $T\gg \delta$ between the paramagnetic and ferromagnetic phases. Although we expect that these conclusions hold also for temperatures $T\lesssim \delta$ we cannot argue it within our approach; a separate analysis is needed.

Similar conclusions about the influence of randomness of the single-particle levels on the Stoner instability can be made for QDs with the Heisenberg exchange~\cite{Future}. Our approach can be extended to the analysis of the effect of level fluctuations on the transverse spin susceptibility and the TDOS for the UH with the Ising exchange.

Our results, in principle, can be checked in QDs made of materials close to the Stoner instability such as Co impurities in a Pd or Pt host, Fe or Mn dissolved in various transition-metal alloys, Ni impurities in a Pd host, and Co in Fe grains, as well as nearly ferromagnetic rare-earth materials~\cite{Exp}. However to test our most interesting results (Eqs. \eqref{eq:Prob:exp} and \eqref{eq:chi:kth}) one needs to explore the regime $(\delta-J_z)/\delta\ll 1/(\pi^2\bm{\beta})$. At present the closest material to the Stoner instability we are aware of, YFe$_2$Zn$_{20}$, has the exchange interaction $J \approx 0.94 \delta$ which is near the boarder of the regime with strong level fluctuations at low temperatures.

We acknowledge useful discussions with Y. Fyodorov, Y. Gefen, A. Ioselevich, A. Shnirman and M. Skvortsov. The research was funded in part by the Russian-Israel scientific research cooperation (RFBR Grant No. 11-02-92470 and IMOST  3-8364), the Council for Grant of the President of Russian Federation (Grant No. MK-4337.2013.2), Dynasty Foundation, RAS Programs ``Quantum mesoscopics and disordered systems'', ``Quantum physics of condensed matter'' and ``Fundamentals of nanotechnology and nanomaterials'', and by Russian Ministry of Education and Science.


\newpage
\begin{widetext}

\centerline{\Large ONLINE SUPPLEMENTAL MATERIAL:}
\vspace{0.2cm}
\centerline{\Large Statistics of Spin Fluctuations in Quantum Dots with Ising Exchange}

\vspace{1cm}
{\small
We present some details on i) the analysis of the correlation function $\overline{V(h_1)V(h_2)}$, ii) the perturbative treatment of the average $\chi_{zz}$, and iii) the calculation of the probability $\Prob\{\ln Z_S>W\}$.
}

\setcounter{equation}{0}

\section{I.\, Correlation function $\overline{V(h_1)V(h_2)}$ \label{Sec1}}

In this section we present a brief derivation of Eqs. (7) and (8) of the paper. The correlation function of the single-particle density of states is given as [S1]
\begin{equation}
\langle \delta\nu_0(E)\delta\nu_0(E+\omega)\rangle = \frac{1}{\delta^2} \left [\delta\left (\frac{\omega}{\delta}\right )- R\left (\frac{\pi \omega}{\delta}\right )\right ] .
\label{App_dnu1}
\end{equation}
Here the function $R(x)$ depends on the statistics of the ensemble of single-particle energies. Using Eq. \eqref{App_dnu1}, the identity $\int_{-\infty}^\infty R(x) dx=\pi$ and the definition of $V(h)$ we obtain
\begin{equation}
\overline{V(h_1)V(h_2)}  = T^2 \int_{-\infty}^\infty \frac{dE d\omega}{\delta^2} R\left (\frac{\pi T \omega}{\delta}\right ) \Bigl [ g(E,h_1) g(E,h_2) - g(E+\omega/2,h_1) g(E-\omega/2,h_2)\Bigr ] ,\label{eqC_app}
\end{equation}
where
\begin{equation}
g(E,h) = \ln \left [ 1+\frac{\sinh^2(\frac{h}{2})}{\cosh^2(\frac{E}{2})}\right ] .
\end{equation}
The function $g(E,h)$ has the following Fourier transform with respect to variable $E$:
\begin{equation}
g(t,h) = \int_{-\infty}^\infty \frac{dE}{2\pi}\, e^{i E t} \, g(E,h) = \frac{1}{2\pi t} \Imag \int_{-\infty}^\infty dE e^{i E t} \tanh\frac{E}{2} \frac{\sinh^2(h/2)}{\sinh^2(h/2)+\cosh^2(E/2)} .
\label{eqgth}
\end{equation}
Since the function $g(E,h)$ is even in $E$, the function $g(t,h)$ is even in $t$. The function under the integral sign in the r.h.s. of Eq. \eqref{eqgth} has poles at $E=\pi (2n+1)i, \pm h + \pi(2m+1) i$ where $n$ and $m$ are integers. Computation of the residues yields
\begin{equation}
g(t,h) = \frac{1}{2\pi t} \Imag 4\pi i \sum_{n\geqslant 0} e^{-\pi(2n+1) t} \left ( 1- \frac{1}{2} e^{-i ht} -\frac{1}{2} e^{i h t}\right ) = \frac{1-\cos(ht)}{t \sinh(\pi t)} .
\end{equation}
Substitution into Eq. \eqref{eqC_app} leads to
\begin{equation}
\overline{V(h_1)V(h_2)}  = 2\pi T^2 \int_{-\infty}^\infty \frac{dt d\omega}{\delta^2} R\left (\frac{\pi T \omega}{\delta}\right ) g(t,h_1) g(t,h_2) \Bigl [ 1- e^{-i\omega t} \Bigr ] .
\end{equation}
At $x\gg 1$ the function $R(x)$ has the following asymptotic behavior [S1]:
\begin{equation}
R(x) = \frac{1}{\bm{\beta} x^2} , \qquad x\gg 1 .
\end{equation}
Recall that $\bm{\beta}=1$ for the orthogonal Wigner-Dyson ensemble, $\bm{\beta}=2$ for the unitary Wigner-Dyson ensemble and $\bm{\beta}=4$ for the simplectic Wigner-Dyson ensemble. Then at $\max\{|h|,T/\delta\}\gg 1$ we find
\begin{equation}
\overline{V(h_1)V(h_2)}  = \frac{4}{\bm{\beta}} \int_{0}^\infty dt\, \frac{[1-\cos(h_1 t)][1-\cos(h_2 t)]}{t \sinh^2(\pi t)} =  \sum_{\sigma=\pm} L(h_1+\sigma h_2)- 2 L(h_1) - 2L(h_2) ,
\end{equation}
where
\begin{equation}
L(h) = \frac{2}{\bm{\beta}} \int_{0}^\infty dt\, \frac{\cos(h t)-1+h^2t^2/2}{t \sinh^2(\pi t)}
\end{equation}
is even in $h$. Next, for $h>0$
\begin{equation}
L^{\prime}(h) = \frac{2}{\bm{\beta}} \int_{0}^\infty dt\, \frac{ht -\sin(h t)}{\sinh^2(\pi t)} = \frac{8}{\bm{\beta}} \int_{0}^\infty dt\, \sum_{n=1}^\infty n [ht -\sin(h t)] e^{-2\pi n t} = \frac{2 h}{\pi^2\bm{\beta}} \left [ \Real \psi\left (1+\frac{i h}{2\pi}\right ) - \psi(1)\right ].
\end{equation}
This is the Eq. (7) of the paper. Using the well-known  asymptotic expressions for the Euler digamma function $\psi(x)$ at small and large values of its argument one arrives at Eq. (8).

\section{II.\, Perturbation theory for the spin susceptibility \label{Sec2}}

In this section we present the derivation of the perturbative results (9) and (10) of the paper. We start the from expansion of the average $\ln Z_S$ to the fourth order in $V$:
\begin{equation}
\overline{\ln Z_S} = \frac{1}{2} \ln \frac{\bar{J}_z}{J_z} - \frac{1}{2} F_{2} - \frac{1}{2} F_{1,1} - \frac{1}{24} F_4 -\frac{1}{8} F_{2,2} - \frac{1}{6} F_{3,1} - \frac{1}{2} F_{2,1,1} - \frac{1}{4} F_{1,1,1,1} + O(V^6) .
\end{equation}
Here we introduced
\begin{equation}
F_{k_1,\dots,k_q} = (-1)^q \int_{-\infty}^\infty \frac{dh_1\dots dh_q}{\pi^{q/2}} \, \exp \left (\sum_{j=1}^q h_j^2\right ) \overline{V^{k_1}(h_1)\dots V^{k_q}(h_q)} .
\end{equation}

\subsection{A.\, Second order in $V$}

The contribution of the second order in $V$ is given by $F_2$ and $F_{1,1}$. We find
\begin{equation}
F_2+F_{1,1} = 2\int_0^\infty \frac{d h}{\sqrt{\pi}} \, e^{-h^2} \left [ 2 L\left (h \sqrt{2\beta \bar{J}_z}\right )- L\left (2h \sqrt{\beta \bar{J}_z}\right ) \right ] .
\label{eq:F2_11}
\end{equation}
It is instructive to compare the second order contribution \eqref{eq:F2_11} with the second order contribution to the variance of $\ln Z_S$:
\begin{equation}
\overline{\bigl (\ln Z_S - \overline{\ln Z_S}\bigr )^2} = F_{1,1} = 4\int_0^\infty \frac{d h}{\sqrt{\pi}} \, e^{-h^2} \left [ L\left (h \sqrt{2\beta \bar{J}_z}\right )- 2 L\left (h \sqrt{\beta \bar{J}_z}\right ) \right ] .
\label{eq:Var}
\end{equation}

In the regime $T\gg \bar{J}_z$ the arguments of $L$ in the r.h.s. of Eqs. \eqref{eq:F2_11} and \eqref{eq:Var} are small. Using the asymptotic expression for $L(h)$ at $|h|\ll 1$, we obtain
\begin{equation}
F_2+F_{1,1} =- \frac{3\zeta(3)}{4\pi^4 \bm{\beta}}\frac{\bar{J}_z^2}{T^2} , \qquad F_{1,1} = \frac{3\zeta(3)}{8\pi^4 \bm{\beta}}\frac{\bar{J}_z^2}{T^2} .
\label{eq:F2+11small}
\end{equation}
The result \eqref{eq:F2+11small} for $F_2+F_{1,1}$ is translated into Eq. (9) of the paper. From Eq. \eqref{eq:F2+11small} we find that
\begin{equation}
\frac{\overline{\bigl (\chi_{zz}-\overline{\chi}_{zz}\bigr)^2}}{\overline{\chi}_{zz}^2} \propto \frac{\bar{J}_z^2}{\pi^2\bm{\beta}T^2} \ll 1, \qquad T\gg \bar{J}_z .
\end{equation}

At low temperatures $T\ll \bar{J}_z$ the asymptotic expression of $L(h)$ for $|h|\gg 1$ must be used in Eq. \eqref{eq:F2_11}. We find
\begin{equation}
F_2+F_{1,1} =- \frac{\ln 2}{\pi^2\bm{\beta}}\frac{\bar{J}_z}{T} , \qquad F_{1,1} = \frac{\ln 2}{\pi^2\bm{\beta}}\frac{\bar{J}_z}{T} .
\label{eq:F2F11}
\end{equation}
From Eq. \eqref{eq:F2+11small} it follows that
\begin{equation}
\frac{\overline{\bigl (\chi_{zz}-\overline{\chi}_{zz}\bigr)^2}}{\overline{\chi}_{zz}^2} \propto \frac{\bar{J}_z}{\pi^2\bm{\beta}T} \ll 1, \qquad \frac{\bar{J}_z }{\pi^2\bm{\beta}}\ll T\ll \bar{J}_z .
\label{eq:Var2}
\end{equation}
In view of the result \eqref{eq:Var2} we can expect that $\ln Z_S$ has a normal distribution with mean
$[\ln (\bar{J_z}/J_z) -F_2-F_{1,1}]/2$ and variance $F_{1,1}$ in the regime ${\bar{J}_z }/{(\pi^2\bm{\beta})}\ll T\ll \bar{J}_z$. For $T=3\delta$ and $J_z/\delta= 0.97$ the CCDF for the normal distribution and the CCDF obtained numerically for the process $V(h)$ are compared in Fig. 1 of the paper. We note that for $T=3\delta$ and $J_z/\delta= 0.97$ numerical integration of  Eqs. \eqref{eq:F2_11} and \eqref{eq:Var} yields $F_2+F_{1,1}\approx -0.09$ and $F_{1,1}\approx 0.06$. These values are still different from the asymptotic estimates \eqref{eq:F2F11}.

\subsection{B.\, Fourth order in $V$}

In the regime $T\gg \bar{J}_z$ the fourth order contributions are proportional to
$(J_z/T)^4$ and therefore negligible. For low temperatures $T\ll \bar{J}_z$ the contributions of the fourth order in $V$ are listed below:
\begin{equation}
F_4 = -3 \int_{-\infty}^\infty \frac{dh}{\sqrt{\pi}} \, e^{-h^2} \Bigl [ \overline{V^2(h)}\Bigr ]^2 = -36 \ln^2 2 \left (\frac{\bar{J}_z}{\pi^2\bm{\beta} T}\right )^2 ,
\end{equation}
\begin{gather}
F_{2,2} = \left [ \int_{-\infty}^\infty \frac{dh}{\sqrt{\pi}} \, e^{-h^2}  \overline{V^2(h)} \right ]^2 +
2  \int_{-\infty}^\infty \frac{dh_1d h_2}{\pi} \, e^{-h_1^2-h^2_2} \Bigl [ \overline{V(h_1)V(h_2)}\Bigr ]^2
=  \left ( 4 \ln^2 2 +  8 b_{2,2} \right ) \left (\frac{\bar{J}_z}{\pi^2\bm{\beta} T}\right )^2 , \\
b_{2,2} = \frac{1}{2} \int_0^{2\pi} \frac{d\phi}{2\pi} \Bigl ( \overline{v(\cos\phi)v(\sin\phi)}\Bigr )^2 \approx 0.35 ,
\end{gather}
\begin{equation}
F_{3,1} = 3 \int_{-\infty}^\infty \frac{dh_1d h_2}{\pi} \, e^{-h_1^2-h^2_2} \overline{V(h_1)V(h_2)}\, \,\overline{V^2(h_2)} = 12 \ln^2 2 \left (\frac{\bar{J}_z}{\pi^2\bm{\beta} T}\right )^2 ,
\end{equation}
\begin{gather}
F_{2,1,1} = - \int_{-\infty}^\infty \frac{dh_1d h_2dh_3}{\pi^{3/2}} \, e^{-h_1^2-h^2_2-h_3^2}\,\,
  \overline{V(h_1)V(h_2)} \,\, \Bigl [ \overline{V^2(h_3)} + 2 \overline{V(h_1)V(h_3)}
\Bigr ]
= - \left (2\ln^2 2 +  2 b_{2,1,1} \right ) \left (\frac{\bar{J}_z}{\pi^2\bm{\beta} T}\right )^2 ,
\\
b_{2,1,1} =\frac{15}{4} \int_0^{2\pi}\frac{d\phi}{4\pi}\int_0^\pi d\theta \, \sin^3\theta \,\,
 \overline{v(\cos\phi)v(\sin\phi)}\,\,\overline{v(\cos\theta)v(\sin\theta\cos\phi)}\approx 0.79 ,
\end{gather}
\begin{equation}
F_{1,1,1,1} = 3 \left [ \int_{-\infty}^\infty \frac{dh}{\sqrt{\pi}} \, e^{-h^2}  \overline{V^2(h)} \right ]^2
= 3 \ln^2 2 \left (\frac{\bar{J}_z}{\pi^2\bm{\beta} T}\right )^2 .
\end{equation}
Summing up, for $T\ll \bar{J}_z$ we obtain
\begin{equation}
\overline{\ln Z_S} = \frac{1}{2} \ln \frac{\bar{J_z}}{J_z} + \frac{\ln 2}{2\pi^2\bm{\beta}}\frac{\bar{J}_z}{T} + \frac{a_2}{4} \left (\frac{\bar{J}_z}{\pi^2\bm{\beta} T}\right )^2 ,
\label{eq:lnZs2d}
\end{equation}
where
\begin{equation}
a_2 = -3\ln^2 2 - 4 b_{2,2} + 4 b_{2,1,1} \approx 0.29 .
\end{equation}
Using Eq. \eqref{eq:lnZs2d} and the definition of the spin susceptibility one can derive Eq. (10) of the paper.

\section{III.\, Complementary cumulative distribution function \label{Sec3}}

In this section we provide some details for the derivation of Eqs. (14) and (15) of the paper. Consider two Gaussian processes,  $Y(t) = -t^2 - z_\gamma v(t)$ and $X(t) = -t^2-(2z_\gamma\sqrt{\ln2}) B(t^2)$ for $t\geqslant 0$. Here $z_\gamma=z/\sqrt{\gamma}$, and $B(t)$ stands for the standard Brownian motion, $\overline{B(t)^2}=2t$. Note that for any time interval $\mathcal{T}$ the sample paths $\{X(t), t\in \mathcal{T}\}$ and $\{Y(t), t\in \mathcal{T}\}$ are bounded a.s., and the following inequalities hold:
\begin{equation}
\begin{split}
\overline{X(t)} & = \overline{Y(t)} ,\\
\overline{X^2(t)} & = \overline{Y^2(t)} ,\\
\overline{[X(t)-X(s)]^2} & \geqslant \overline{[Y(t)-Y(s)]^2}
\end{split}
\label{eq:SlepianIneq}
\end{equation}
for all $t, s \in \mathcal{T}$. Indeed, the first two conditions are trivially satisfied, while the last one follows from an easily verifiable inequality
\begin{equation}
\overline{[v(1/2+r)-v(1/2-r)]^2} \leqslant 8 r \ln 2
\end{equation}
($|r| \leqslant 1/2$). Hence we are in position to apply the Slepian's inequality [S2] to the pair $(X(t),Y(t))$ and claim that for all real $w$
\begin{equation}
\Prob\{\max\limits_{t\in \mathcal{T}} X(t)>w\}\geqslant \Prob\{\max\limits_{t\in \mathcal{T}} Y(t) > w\} .
\label{eq:SlepianIneq1}
\end{equation}
A well-known result for the Brownian motion with linear drift (see, e.g., [S3]) reads ($w>0$)
\begin{equation}
\Prob\{\max\limits_{t\geqslant 0} X(t)>w\} = \exp\left (- \frac{w}{2z_\gamma^2\ln 2}\right ) .
\label{eq:Brownian}
\end{equation}
Combining Eq. \eqref{eq:SlepianIneq1} and Eq. \eqref{eq:Brownian}, we obtain the inequality (14) of the paper.

The above reasoning has the advantages of being rigorous and self-contained, but it provides only an upper bound for the CCDF tail; one can in fact go further. As it can be seen from rescaling the time axis, the probability that the maximum of $Y(t)=-t^2-z_\gamma v(t)$ exceeds $w$ equals the probability that the maximum of $\widetilde Y(s)=v(s)/(1+s^2)$ defined on $s\geqslant 0$ exceeds $w^{1/2}/z_\gamma$. From the results of H\"usler and Piterbarg [S4] it follows that the large-$w$ tail of $\Prob\{\max\limits_{t\geqslant 0} Y(t) > w\}$ is determined by a small vicinity of the point $s^*=1$ where the variance of $\widetilde Y(s)$ attains its maximum $\ln 2$. Furthermore, should we have a finite limit
\begin{equation}
\lim\limits_{s, t \to s_*}
\frac{\overline{[\widetilde Y(s) - \widetilde Y(t)]^2}}
{K^2(s-t)} >0
\label{eq:Kdef}
\end{equation}
for some function $K(x)$ regularly varying at $0$ with index $\alpha\in(0,1)$, the precise asymptotics would read
\begin{equation}
\Prob\{\max\limits_{t\geqslant 0} Y(t) > W\} \sim {\rm const}(\alpha)\cdot \frac{ (z^2/W)^{-1}}{K^{-1}\bigl (\sqrt{z^2/W}\bigr)} \exp \left [-\frac{W}{2z^2\ln 2}\right ] , \quad W/z^2 \gg 1
\label{eq:ProbTail}
\end{equation}
where $K^{-1}(x)$ stands for the functional inverse of $K(x)$. In our case $\overline{[v(h+u)-v(h)]^2} \to - 2u^2\ln |u| +O(u^2)$ translates into $K(x)=x\sqrt{\ln(1/x)}$ which is regularly varying with index $\alpha=1$ (recall that a function $f(x)$ is regular varying at~0 with index $\alpha$ if $\lim_{t\to 0} f(at)/f(t)=a^{\alpha}$ for any $a>0$). The result of Ref. [S4] is therefore not directly applicable, but we believe this to be a technicality. In analogy with a similar situation for fractional Brownian motion, we expect the asymptotics \eqref{eq:ProbTail} to hold with only the $W$-independent factor ${\rm const}(\alpha)$ modified. Note that the exponential part can be tracked to be the tail of a normal distribution with variance $\ln 2$ taken at $W^{1/2}/z$, and that it had been correctly reproduced by our initial estimate.

For $z\gg 1$ the saddle-point approximation in Eq. (6) of the paper becomes exact and the statistics of $\ln Z_S$ reduces to the statistics of maxima of $Y(h)$ directly. For $x\ll 1$ the inverse of $K(x)$ is $K^{-1}(x) \approx x/\sqrt{\ln 1/x^2}$ (with logarithmic accuracy), and we arrive at Eq. (15) of the paper.

To illustrate that result we approximate the Gaussian process $v(h)$ by a degenerate one $\tilde{v}(h) = \xi |h|$, where $\xi$ is the Gaussian random variable with zero mean $\overline{\xi}=0$ and variance $\overline{\xi^2}=4\ln 2$. Then for a given $\xi$ we find (see Eq. (6) of the paper)
\begin{equation}
Z_S \simeq \mathcal{Z}_S(\xi) \equiv \left (\frac{\bar{J}_z}{J_z}\right )^{1/2} e^{z^2 \xi^2/4} \Bigl [1 - \erf(z \xi/2) \Bigr ] .
\label{eq:ZS:xi}
\end{equation}
Large values of $Z_S$ correspond to large negative values of $\xi$ such that $\ln Z_S \approx z^2 \xi^2/4$. On the other hand, since $Z_S \geqslant 1$ Eq. \eqref{eq:ZS:xi} limits variable $\xi$ to be in the interval $-\infty < \xi < \xi_0$ where $\xi_0$ is given as the solution of equation $\mathcal{Z}_S(\xi_0)=1$. It can be estimated as $\xi_0 \approx (2/z)\sqrt{J_z/(\pi \bar{J}_z)}=\sqrt{\pi \bm{\beta}T/J_z} \gg 1$. Then, within the degenerate case, we find the following result for the complementary cumulative distribution function:
\begin{equation}
\mathcal{P}_{\rm deg}(W) = \frac{1+\erf\bigl (\xi_W/\sqrt{8\ln 2}\bigr )}{1+\erf\bigl (\xi_0/\sqrt{8\ln 2}\bigr )},
\label{eq:PW:bal}
\end{equation}
where $\xi_W$ is the solution of the equation $\ln \mathcal{Z}_S(\xi_W)=W$. At $z\gg 1$, for $W\gg z^2$ one finds $\xi_W\approx 2 \sqrt{W}/z$ and, hence, that the tail of the function $\mathcal{P}_{\rm deg}(W)$ is given by Eq. (15) of the paper without the logarithm in the pre-exponent. The function $\mathcal{P}_{\rm deg}(W)$ for $T=3\delta$ and $J_z/\delta=0.9997$ is shown as the red dashed curve in Fig. 1 of the paper.

{\small
\begin{itemize}

\item[[S1\!\!]] M.L. Mehta, {\it Random Matrices} (Boston: Academic) (1991).

\item[[S2\!\!]] R.J. Adler, {\it An Introduction to continuity, extrema, and related topics for general Gaussian processes}, (Hayward, California, 1990).

\item[[S3\!\!]] S. Asmussen, H. Albrecher, {\it Ruin Probabilities}
 (World Scientific, 2010).

\item[[S4\!\!]] J. H\"usler, V. Piterbarg,
Stoch. Proc. Appl. {\bf 83}, 257 (1999).

\end{itemize}
}

\end{widetext}

\end{document}